# One Sample Diffusion Model in Projection Domain for Low-Dose CT Imaging

Bin Huang, Liu Zhang, Shiyu Lu, Boyu Lin, Weiwen Wu, *Member, IEEE*
Qiegen Liu, *Senior Member, IEEE*

*Abstract*—Low-dose computed tomography (CT) plays a significant role in reducing the radiation risk in clinical applications. However, lowering the radiation dose will significantly degrade the image quality. With the rapid development and wide application of deep learning, it has brought new directions for the development of low-dose CT imaging algorithms. Therefore, we propose a fully unsupervised one sample diffusion model (OSDM) in projection domain for low-dose CT reconstruction. To extract sufficient prior information from single sample, the Hankel matrix formulation is employed. Besides, the penalized weighted least-squares and total variation are introduced to achieve superior image quality. Specifically, we first train a score-based generative model on one sinogram by extracting a great number of tensors from the structural-Hankel matrix as the network input to capture prior distribution. Then, at the inference stage, the stochastic differential equation solver and data-consistency step are performed iteratively to obtain the sinogram data. Finally, the final image is obtained through the filtered back-projection algorithm. The reconstructed results are approaching to the normal-dose counterparts. The results prove that OSDM is practical and effective model for reducing the artifacts and preserving the image quality.

## I. INTRODUCTION

Over the last few decades, X-ray computed tomography (CT) techniques have been widely employed in clinical for diagnosis and intervention, including imaging, image-guided needle biopsy, image-guided intervention, and radiotherapy with noticeable benefits [1, 2]. However, the associated x-ray radiation dose which may potentially induce lifetime risk of cancers has attracted wide public attention [3]. In general, an effective approach for reducing radiation dose is to lower the x-ray exposure [4-6] by adjusting the tube current, which can lead to the sharp artifacts and noise. Thereby, the low-dose CT image is always accompanied with strong noise and features missing [7].

Low-dose CT reconstruction is one of classic inverse problems. Many algorithms have been proposed, they can be mainly grouped into three categories: image post-processing methods [8, 9], sinogram domain methods [10-12] and iterative reconstruction methods [13-15]. For image post-processing methods, they can directly process low quality images as the input and ground truth as the outputs without raw projections data. Zheng *et al.* [8] exploited sparse representation and image decomposition theory to separate low-dose CT images from noise. Based on the popular idea of sparse representation, Chen *et al.* [9] presented a patch-based dictionary learning approach for effective suppression of both mottled noise and streak artifacts. Nevertheless, it is difficult to remove severe streak artifacts and accurate recover image details and features without projection data. The sinogram domain based reconstruction methods is beneficial to solve this problem. Yin *et al.* [10] put forward a domain progressive 3D residual convolution network. Humphries *et al.* [11] studied the performance of a simple CNN-based approach to low-dose CT employing both low intensity and sparse view scans. Ghani *et al.* [12] applied deep-learning to denoise the original sinograms. The iterative reconstruction concentrated to solving the low-dose CT problem iteratively by extracting prior information on target images [3, 13-14]. Various priors were developed, the most well-known one is the total variation (TV) based reconstructions [15-17]. For that, Liu *et al.* [15] defined a local TV and improved wavelet residual convolutional neural network denoising model. Furthermore, Deng *et al.* [16] proposed a denoising model for projection data utilizing noise level weighted TV regularization term. Sagheer *et al.* [17] took tensor TV into consideration, developing a low-rank approximation based approach to improve global smoothness. Meanwhile, the iterative reconstruction methods heavily hurt computational costs.

Recently, diffusion models open a new door to address image processing tasks with a fresh perspective [18-20]. For instance, Lyu *et al.* [21] proposed a conditional DDPM method for low-dose CT reconstruction. In addition, by incorporating the forward and backward diffusion processes into the stochastic differential equation (SDE) framework, Song *et al.* [22] designed the score-based generative model which consists of the forward and backward diffusion processes in the SDE framework. The model has been widely applied in clinical scenarios and achieved excellent results [23-25]. In the meanwhile, compared with GAN, diffusion models achieve great performance.

Due to the scarcity and infeasible of data, especially clinical data, more and more researchers focus on one-shot and few-shot reconstruction studies. A popular solution is to expand the data by changing the scale. To capture internal statistical properties at different scales, large images are successively subsampled for attaining sufficient small image patches as the training set. Inspired by the data redundancy and structural reproducibility, the Hankel matrix is a common strategy in image denoising [26], artifact removal [27] and deconvolution [28]. Based on the characteristics of Hankel matrix, Wang *et al.* [29] designed an encoder-decoder network from the learning-based Hankel matrix decomposition. Supporting that the Hankel matrix is large

This work was supported by National Natural Science Foundation of China (61871206). (Corresponding author: Weiwen Wu and Qiegen Liu.)
This work did not involve human subjects or animals in its research.
B. Huang, L. Zhang, S. Lu, B. Lin, and Q. Liu are with School of Information Engineering, Nanchang University, Nanchang 330031, China. ({huangbing, zhangliu, lushiyu, linboyu}@email.ncu.edu.cn, {liuqiegen}@ncu.edu.cn)
W. Wu is with the School of Biomedical Engineering, Sun Yat-Sen University, Shenzhen, Guangdong, China (wuweiw7@mail.sysu.edu.cn).

enough, then it maybe contains enough internal statistics information for image reconstruction. Thereby, it is possible that low-dose images can be reconstructed into normal-dose images by employing one-shot or few-shot samples via low rank Hankel matrix formation. Meanwhile, although studies on one sample or few samples of natural images have emerged, there are seldom one sample or few samples on medical images [30-32].

In this study, we propose a one sample based diffusion model method in the projection domain for low-dose CT reconstruction, namely OSDM. Specifically, we combine low-rank structural-Hankel matrix with the diffusion model to generate the ideal sinogram from the low-dose projection data. Besides, penalized weighted least-squares (PWLS) and TV are introduced to achieve superior image quality. Our main idea is to learn the prior distribution from normal-dose sinogram by the diffusion model and infer the lost information from low-dose sinogram. Unlike previous supervised learning methods, our unsupervised method does not require retraining on low-dose/normal-dose CT image pairs when the dosage of projection changes. Notably, only one data is required for training.

What's more, training the model in projection domain is more conducive to the generalization of the model. For example, there are huge differences between brain and lung in image domain data, while it is similar in projection domain. Nevertheless, the generalization of model can be improved effectively by the projection diffusion models. Thus, we developed our OSDM reconstruction method in the projection domain rather than image domain.

The theoretical and practical contributions of this work are summarized as follows:
- To alleviate the shortage of medical samples, low-rank Hankel matrix is constructed to infer the internal statistics within the CT sinogram. Only one sample is required to extract sufficient prior information.
- Employing a score-based diffusion model for projection domain interpolation. As training sets or test sets turn, the proposed method maintains great generalization without retraining models.

The rest of the manuscript is organized as follows. Relevant background on score-based diffusion models and the construction process of Hankel matrix are demonstrated in Section II. Detailed procedure and algorithm of the proposed method is presented in Section III. Experimental results and specifications about the implementation and experiments are given in Section IV. At last, the conclusion is drawn in Section V.

## II. PRELIMINARY

### A. Reconstruction of Low-dose CT Images

Low-dose CT reconstruction is a classic inverse problem, which seeks to reconstruct the fuzzy part from existing low-dose CT images. In particular, assuming that $x \in \mathbb{R}^N$ is the degraded sinogram, the forward formulation of sinogram reconstruction problem could be given by:
$$y = x + n \quad (1)$$
where $n \in \mathbb{R}^M$ represents for additive noise. $y$ denotes a low-dose sinogram. Note that the inverse problem amounts to reconstructing $x$ from $y$.

For avoiding ill-posed, the problem of reconstructing CT sinograms is formulated as a constrained optimization equation as follows:
$$\min_x \|x - y\|_2^2 + \mu R(x) \quad (2)$$
where $\|x-y\|_2^2$ is the data fidelity term. $R(x)$ denotes the regulation prior knowledge term, which is chosen to be a TV semi-norm. $\|\bullet\|_2^2$ represents the $l_2$ norm. Besides, $\mu$ is the factor to keep a good balance between the data-consistency (DC) term and the regularization term.

Taking the TV term into consideration, this optimization process differentiates infinite solutions to the Eq. (2) and picks out the best one with desired image properties as the reconstructed sinogram. Generally, the TV term is defined as:
$$R(x) = \|x\|_{TV}^2 = \int_\Omega |\nabla x| dx \quad (3)$$
where $\Omega$ is a bounded domain. $\nabla x$ represents the gradient of a sinogram $x$. The TV term has been shown to be robust to remove noise and artifacts in the reconstructed sinogram [3, 33].

### B. Construction of Hankel Matrix

The Hankel matrix can take advantage of data redundancy to extract internal statistics within low-dose CT sinograms. Based on the idea that a low-rank matrix could express projection data, the new data formulation is designed here to manifest inner relationship in sinograms.

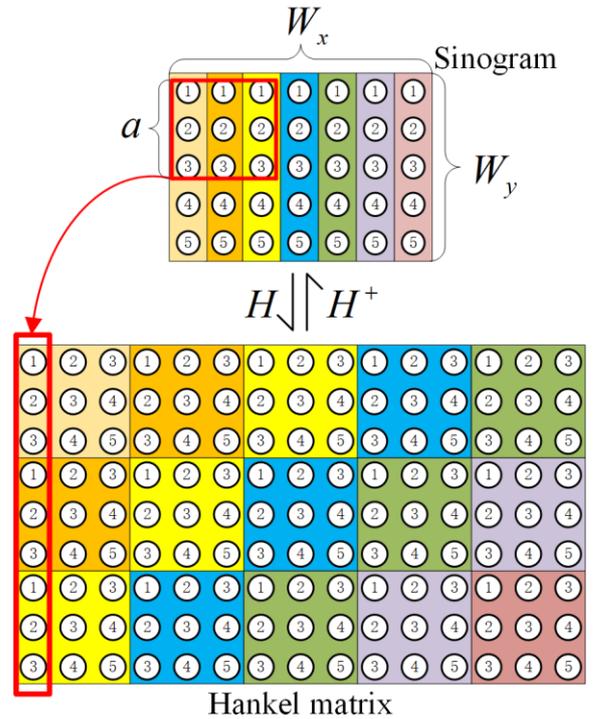

**Fig. 1.** Constructing a new data formulation from one sinogram ( $H$ ) and vice versa ( $H^+$ ).

As illustrated in Fig. 1, the sinogram can be transformed to a new data formulation by sliding a window with the size of $a \times a$ across entire sinogram. More carefully, the size of sinogram is $W_x \times W_y$. After the transformation $H$, the new data formulation with a size of $a^2 \times (W_x - a + 1)(W_y - a + 1)$ is generated. Fig. 1 demonstrates the case of $W_x = 7$, $W_y = 5$ and $a = 3$. Besides, the step length of sliding window is set to be 1.

Individual blocks data in the projection domain are vec-torized as columns in the new data formulation. The linear operator $H$ is defined as generating new data formulation from the sinogram concatenated in a vector form:

$$H := R^{W_x \times W_y} \to R^{a^2 \times (W_x-a+1)(W_y-a+1)} \quad (4)$$

When reversely forming a sinogram dataset from the new data formulation, multiple antidiagonal entries are averaged and stored in the projection domain. Hence, the reverse operator $H^+$ generates the corresponding projection dataset from the new data formulation, and is given as follows:

$$H^+ := R^{a^2 \times (W_x-a+1)(W_y-a+1)} \to R^{W_x \times W_y} \quad (5)$$

where + stands for a pseudo-inverse operator. It is equivalent to averaging the anti-diagonal elements and placing them in the appropriate locations.

*C. Score-based SDE*

The great success of diffusion models, especially score-based SDE in creating realistic and variable image samples has aroused widespread concern [22]. Score-based SDE comprises of the forward process and the reverse-time process.

Given a continuous diffusion process $\{x(t)\}_{t=0}^T$ with $x(t) \in \mathbb{R}^N$, which is indexed by $t \in [0,T]$ as the progression time variable. $N$ stands for the sinogram dimension. The diffusion process can be formulated as the solution to the following SDE:

$$dx = f(x,t)dt + g(t)dw \quad (6)$$

where $f(x,t) \in \mathbb{R}^N$ and $g(t) \in \mathbb{R}$ correspond to the drift coefficient and diffusion coefficient, respectively. $dt$ corresponds to an infinitesimal positive time step and $w \in \mathbb{R}^N$ induces the Brownian motion.

Via reversing the above process, samples can be attained. Notably, the reverse-time SDE is also a diffusion process, which could be expressed as follows:

$$dx = [f(x,t) - g(t)^2 \nabla_x \log p_t(x)]dt + g(t)d\bar{w} \quad (7)$$

where $\bar{w}$ is a standard Wiener process when time flows backwards from $T$ to $0$, and $dt$ is an infinitesimal negative time step.

## III. PROPOSED METHOD

*A. Low-dose CT Imaging Model*

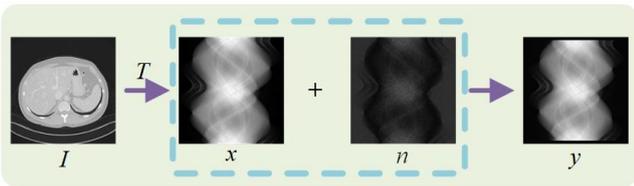

**Fig. 2.** Linear measurement process for low-dose CT.

As shown in Fig. 2, the linear measurement process for low-dose CT imaging is visualized. Intuitively, $n$ can be defined as the low-dose noise on the sinogram. $I$ represents CT images. $T(\cdot)$ corresponds to the Radon transform. If an ideal sinogram $x$ of size $768 \times 768$ is measured in the presence of $n$, the CT reconstruction problem can be formulated as solving following equation:

$$y = T(I) + n = x + n \quad (8)$$

where $y$ is the low-dose CT projection data of size $768 \times 768$. In addition, $T^{-1}(\cdot)$ can be implemented with the inverse Radon transform.

According to abundant experiments of real projection data, sinograms scanned by decreased current intensity can be approximately viewed as a result of ideal projection data contaminated by additive noise [34-38]. The additive noise is approximately subject to Poisson distribution. Especially, a Poisson model for the intensity measurement is presented by:

$$L_i \sim Poisson\{a_i e^{-[x]_i} + r_i\}, \quad i = 1, \cdots, N_m \quad (9)$$

where $L_i$ denotes the number of transmitted photons, $a_i$ is the X-ray source intensity of the $i$-th ray, and $r_i$ indicates the background contributions of scatter and electrical noise. $x$ corresponds a vector for the representation of attenuation coefficients with units of inverse length, $N_m$ stands for the number of measurements and $N_v$ stands for the number of image voxels.

Via taking the logarithm operation, the measurement data is transformed to a weighted Gaussian formulation [39]:

$$y_i \sim N([x]_i, \bar{L}_i / (\bar{L}_i - r_i)^2) \quad (10)$$

where $\bar{L}_i = E[L_i]$. In fact, the low-dose CT problem is expressed as a typical inverse problem $y = x + n$. In order to address uncertain ill-posed problems, the posterior distribution $p(x|y)$ is introduced by the theory of Bayesian inversion [40]. Thereby, the inverse problem is transformed to a problem conditioned on the measurements $y$.

*B. Data Preprocessing*

During the training phase, we need to conduct data processing on the structural-Hankel matrix formulation before entering the network. The process of training model can be visualized in Fig. 3.

For storing the internal statistics within normal-dose projection data, the Hankel matrix with the size of $579121 \times 64$ is constructed by sliding a window with the size of $8 \times 8$ on initial normal-dose projection data. After Hankel transformation, the same information in initial normal-dose projection data appears in diverse locations of the Hankel matrix. Thereby, the redundancy of the Hankel matrix will be employed to capture the internal statistics within normal-dose projection data. Crucially, here only one normal-dose sinogram is needed.

Suppose the initial normal-dose projection data is denoted as $x$, the construction process of Hankel matrix $H$ is given by:

$$H_s = H(x) \quad (11)$$

where $H(\cdot)$ corresponds to the Hankel transformation and $H_s$ represents constructed Hankel matrix.

The Hankel matrix is cropped into high-dimensional data owning the size of $64 \times 64 \times 9048$ to better employ the prior learning knowledge. Notably, a large number of small patches are spilt from the constructed Hankel matrix in a random way, which is shown in Fig. 3. The specific process can be expressed by:

$$X = S(H_s) \quad (12)$$

where $S(\cdot)$ represents the random split operation. $X$ stands for the high-dimensional tensor. Plenty of tensors will be constructed through random split operation and treated as the network input. This data amplification method can enhance the training set to get sufficient prior knowledge.

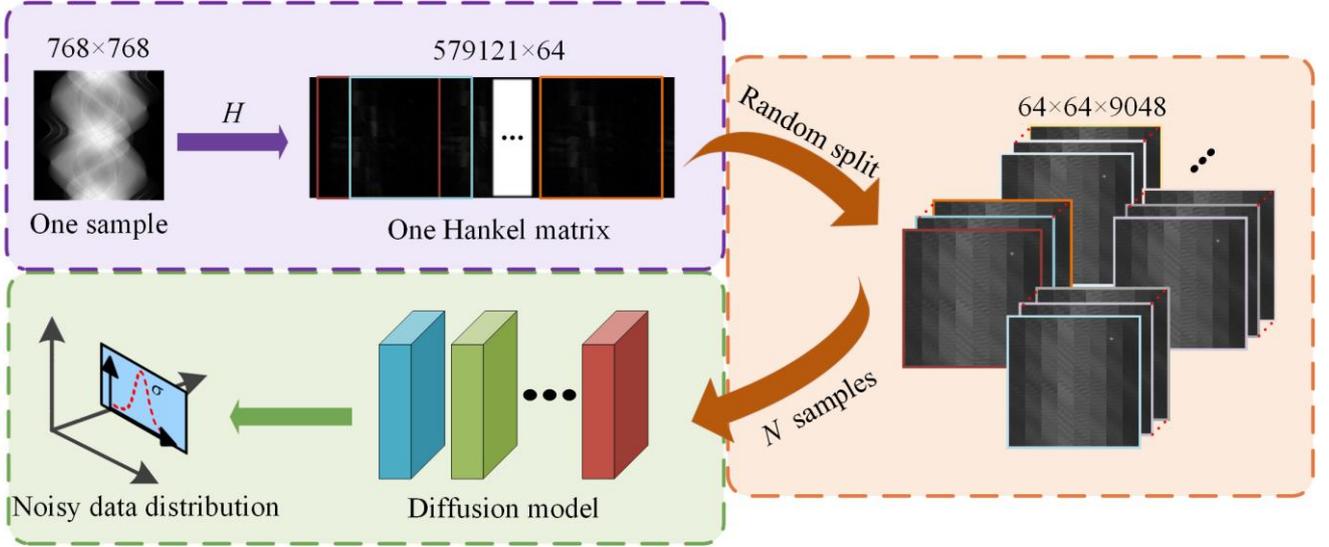

**Fig. 3.** The pipeline of OSDM training process. Before the prior learning of the score-based network, one normal-dose sinogram is injected by multi-scale noise gradually. Then, we construct the low-rank Hankel through Hankel transformation to obtain sufficient redundant information. Finally, the Hankel matrix is spilt into multi-tensors (i.e., *N* samples) in a random order and the network learns the gradient distribution via denoising score matching.

Based on the description in section II. C, the score-based model learns prior distribution via leveraging an SDE. The forward SDE smoothly transforms a complex data distribution to a known prior distribution by slowly injecting noise. In Fig. 4, we provide a demonstration of above two processes.

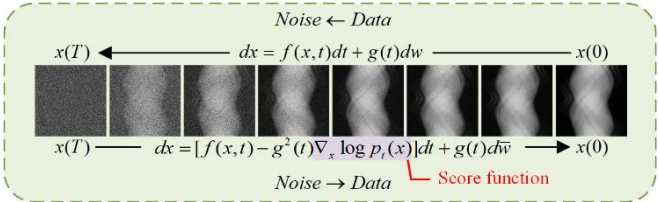

**Fig. 4.** The perturbed data by noise is smoothed with following the trajectory of an SDE. By estimating the score function $\nabla_x \log p_t(x)$ with SDE, it is possible to approximate the reverse SDE and then solve it to generate sinogram samples from noise.

During the training phase, the peak performance of network can be reached by optimizing the parameters $\theta^*$ of the score-based network. The objective function can be described as:

$$\theta^* = \arg\min_\theta \mathbb{E}_t \{\lambda(t)\mathbb{E}_{x(0)}\mathbb{E}_{x(t)|x(0)}[ \\ \|s_\theta(x(t),t) - \nabla_{x(t)} \log p_t(x(t)|x(0))\|_2^2]\} \quad (13)$$

where $\gamma:[0,T] \to \mathbb{R}_{>0}$ is a positive weighting function and $t$ is uniformly sampled over $[0,T]$. $p_t(x(t)|x(0))$ is the Gaussian perturbation kernel centered at $x(0)$. Once the network satisfies $s_\theta(x(t),t) \simeq \nabla_x \log p_t(x)$, which means that $\nabla_x \log p_t(x)$ is known for all $t$ by solving $s_\theta(x(t),t)$.

### C. OSDM: Iterative Reconstruction

The iterative reconstruction process is demonstrated in Fig. 5. At first, the sinogram data $x$ is transformed from the CT image $I$ via $T^{-1}$ operation which corresponds to forward projection (FP):

$$x = T^{-1}(I) \quad (14)$$

To suppress noise and obtain richer information, the prior distribution $p_t(x)$ of sinogram data is estimated by employing the score-based diffusion model. In the forward diffusion process, rather than perturbing data with a finite number of noise distributions, a continuous distribution over time is considered. By reversing SDE, we can convert random noise into data for sampling. As suggested in [41], the predictor-corrector (PC) sampling is introduced at samples updating step in this paper. As part of PC sampling, the predictor is considered as a numerical solver for the reverse-time SDE. When the reverse-time SDE process ends, the samples are generated based on the prior distribution, which can be discretized as follows:

$$x^i \leftarrow x^{i+1} + (\sigma_{i+1}^2 - \sigma_i^2)s_\theta(x^{i+1},\sigma_{i+1}) + \sqrt{\sigma_{i+1}^2 - \sigma_i^2}z \quad (15)$$
$$i = N-1,\cdots,0$$

where $z \sim N(0,1)$, $x(0) \sim p_0$, and $\sigma_0 = 0$ is chosen to simplify the notation. The above formulation is repeated for $i = N-1,\cdots,0$. After adding the conditional constraints to Eq. (17), the formula could be rewritten as follows:

$$x^i = x^{i+1} + (\sigma_{i+1}^2 - \sigma_i^2)\nabla_x[\log p_t(y|x^{i+1}) + \log p_t(x_{LR}^{i+1}) \\ + \log p_t(x_{TV}^{i+1})] + \sqrt{\sigma_{i+1}^2 - \sigma_i^2}z \quad (16)$$

In the above equation, $\log p_t(y|x)$ stems from sinogram data knowledge. $\log p_t(x_{LR})$ is derived from low rank (LR) data knowledge and $\log p_t(x_{TV})$ comes from TV data knowledge.

Then, the updated sample $x$ is transformed to a new data formulation through $H(\bullet)$ Hankel Transform (HT) operation:

$$H_p^i \leftarrow H(x^i) \quad (17)$$

**LR Step.** For easy manipulation and ready analyzation, the Hankel matrix is decomposed into the form (18) via the singular value decomposition (SVD) operation.

$$[U\Delta V^T] = svd(H^i) \quad (18)$$

where $U$ is an orthogonal matrix, $\Delta$ is a diagonal matrix with non-negative diagonal elements, while $V$ is an orthogonal matrix. Precisely, $H$ is a 579121×64 matrix of rank $L$ (i.e., $H$ contains $L$ singular values that are not zero) while $U_{[k]}$, $V_{[k]}$, and $\Delta_{[k]}$ represents the first $K$ columns of $U$, $V$, and $\Delta$, respectively:

$$U_{[k]} = [u_1, \cdots, u_k, \cdots, u_K]$$
$$V_{[k]} = [v_1, \cdots, v_k, \cdots, v_K] \quad (19)$$
$$\Delta_{[k]} = [\delta_1, \cdots, \delta_k, \cdots, \delta_K]$$

$H_{[k]}$ represents the matrix $H$ that reconstructed from the first $K$ eigenvectors. The hard-threshold (hard-THR) singular value process can be expressed as:

$$H_{[k]}^i = U_{[k]} \Delta_{[k]} V_{[k]}^T \quad (20)$$

In particular, the SVD gives the best approximation with smaller rank. This way is extremely useful for ill-defined linear problem with almost degenerate matrices.

After low-rank process, Hankel matrix is transformed to sinogram again through $H^+(\bullet)$, where $H^+(\bullet)$ represents inverse Hankel transform (IHT) operation:

$$x^i \leftarrow H^+(H_{[k]}^i) \quad (21)$$

***TV Step.*** The TV minimization operation is also conducted for removing noise and artifacts. Suppose $\Delta x = \|x - x^i\|$, and TV minimization can be stated as follows:

$$TV(x^i) = x^{i+1} - \alpha \times \Delta x \times \frac{\nabla \|x^i\|_{TV}}{\|\nabla \|x^i\|_{TV}\|} \quad (22)$$

where $\alpha$ is the length of each gradient-descent step.

***DC Step.*** It's possible to improve anti-noise properties by the method of adding statistical characteristics of projection data to object function [34-38, 42]. And, there is a statistical approach for sinogram denoising, which is finding an optimal valuation of the sinogram from the noisy sinogram by PWLS method. The PWLS prior is incorporated into regularized objective function. The regularized objective function can be expressed as:

$$x^i = \arg\min_x [\|y - x^{i+1}\|_W^2 + \lambda_1 \|x^{i+1} - H^+(H_{[K]}^{i+1})\|_2^2 + \lambda_2 \|x^{i+1}\|_{TV}^2] \quad (23)$$

where hyperparameter $\lambda_1$, $\lambda_2$ balances the trade-off among the terms of PWLS, LR and TV. $i = N-1, \cdots, 0$ denotes the iteration of outer loops. Specifically, the standard PWLS can be described as follows:

$$x^i = \arg\min_x [(x^{i+1} - y)^T W (x^{i+1} - y) + \mu R(x^{i+1})] \quad (24)$$

where superscript $T$ represents transposing operation. Eq. (24) can be further solved as:

$$x^i = \frac{W(y - x^{i+1}) + \mu R'(x^{i+1})}{W + \mu} \quad (25)$$

In order to decrease the influence of noise on the effect of reconstruction, the scale coefficient $\eta$ for system calibration is set to 22000.

$$W = diag\{w_i\} = diag\left\{\frac{1}{\sigma_{x_i}^2}\right\} = diag\left\{\frac{1}{l_1 \exp(x_i/\eta)}\right\} \quad (26)$$

When it comes to the corrector, it refers to the Langevin dynamics via transforming any initial sample $x(t)$ to the final sample $x(0)$ with the following procedure:

$$x^{i,j} \leftarrow x^{i,j-1} + \varepsilon_i s_\theta(x^{i,j-1}; \sigma_i) + \sqrt{2\varepsilon_i} z \quad (27)$$
$$j = 1, 2, \cdots, M, \quad i = N-1, \cdots, 0$$

where $\varepsilon_i > 0$ is the step size, and $z \sim N(0,1)$ refers to a standard normal distribution. The above formulation is repeated for $j = 1, 2, \cdots, M$, $i = N-1, \cdots, 0$. The theory of Langevin dynamics guarantees that when $M \to \infty$ and $\varepsilon_i \to 0$, $x^i$ is a sample from $p_t(x)$ under designated conditions.

The above-mentioned sampling is not directly from $p(x)$, but from the posterior distribution $p(x|y)$ by employing SDE in Section II. An intuitive solution is that the DC in Eq. (8) can be considered as a conditional term which can be incorporated into the sampling procedure of Eq. (27), and it yields:

$$x^{i,j} = x^{i,j-1} + \varepsilon_i s_\theta(x^{i,j-1}; \sigma_i) + \sqrt{2\varepsilon_i} z \quad (28)$$

Once we obtain the reconstructed projection $x$, the final image $\tilde{I}$ is obtained:

$$\tilde{I} = T(x) \quad (29)$$

where $T(\bullet)$ means the filtered back-projection (FBP).

| Algorithm 1: Training and reconstruction stages |
|---|
| **Training stage** |
| **Dataset:** One sample $x$ in the projection domain |
| 1: **Repeat** |
| 2: $\quad x \sim p(x)$, $t \sim \mathcal{U}([0,T])$, $\varepsilon \sim \mathcal{N}(0,I)$ |
| 3: $\quad x(t) = x(0) + \varepsilon \sigma(t)$ |
| 4: $\quad$ Take a gradient descent step on $\nabla_\theta \|s_\theta(x(t),t) + \varepsilon\|_2^2$ |
| 5: **Until** converged |
| 6: Trained OSDM |
| **Reconstruction stage** |
| **Setting:** $s_\theta, N, M, \sigma, \varepsilon$ |
| 1: Initial data $x = T^{-1}(I)$ **(FP)** |
| 2: $x^N \sim \mathcal{N}(0, \sigma_{max}^2 I)$ |
| 3: For $i = N-1$ to $0$ do **(Outer loop)** |
| 4: $\quad x^i \leftarrow Predictor(x^{i+1}, \sigma_i, \sigma_{i+1})$ |
| 5: $\quad H_p^i \leftarrow H(x^i)$ **(HT)** |
| 6: $\quad [U \Delta V^T] = svd(H_p^i)$ **(SVD)** |
| 7: $\quad H_{[k]}^i = U_{[k]} \Delta_{[k]} V_{[k]}^T$ **(hard-THR)** |
| 8: $\quad x^i \leftarrow H^+(H_{[k]}^i)$ **(IHT)** |
| 9: $\quad x^i = \dfrac{W(y - x^{i+1}) + \mu R'(x^{i+1})}{W + \mu}$ **(PWLS)** |
| 10: $\quad x^i = TV(x^i)$ **(TV)** |
| 11: $\quad$ For $j = 1$ to $M$ do **(Inner loop)** |
| 12: $\quad\quad x^{i,j} \leftarrow Corrector(x^{i,j-1}, \sigma_i, \varepsilon_i)$ |
| 13: $\quad\quad$ Repeat from step 5 to step 10 |
| 14: $\quad$ End for |
| 15: End for |
| 16: Final image $\tilde{I} = T(x)$ **(FBP)** |
| 17: Return $\tilde{I}$ |

Furthermore, **Algorithm 1** demonstrates training and reconstruction processes of OSDM in detail. During the training passe, only one sinogram sample is required. The whole OSDM reconstruction process encompasses two loops. In the outer loop, predictor is performed with trained network. In the inner loop, corrector is performed for correction. The predictor and the corrector work together as a whole to generate final samples. Besides, the data prior items and data fidelity items of the prediction and correction process are updated in each loop.

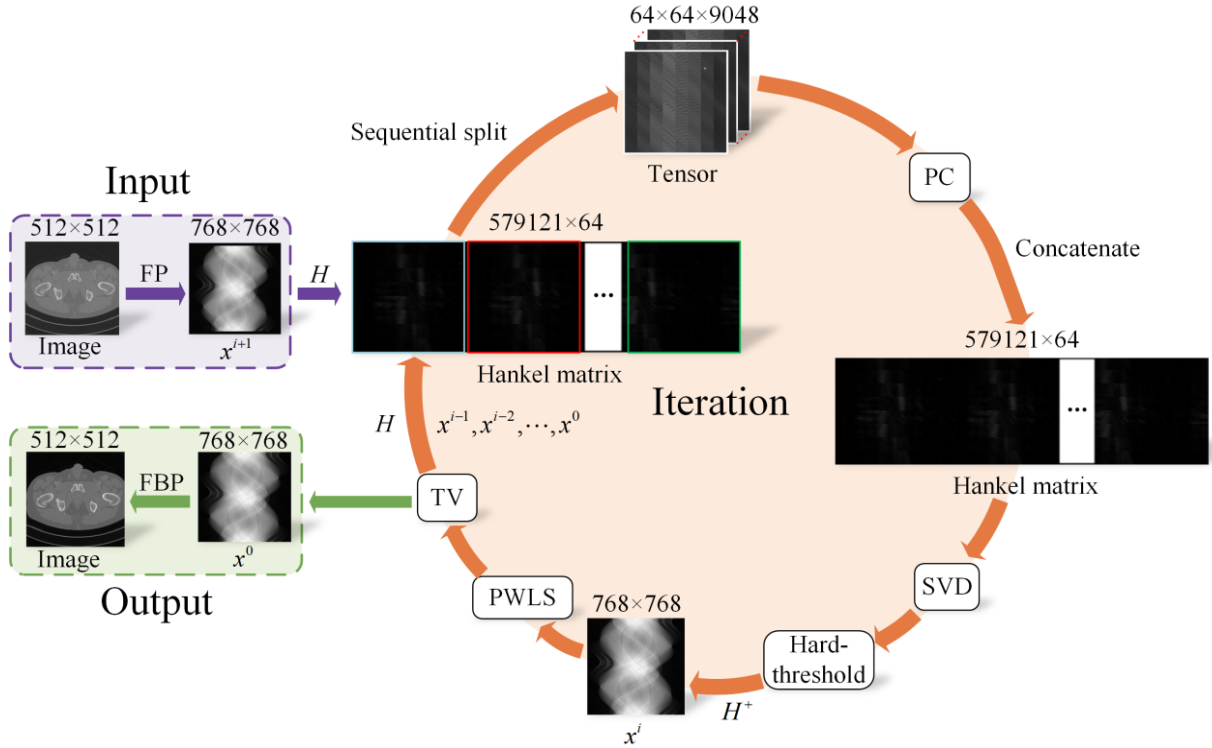

**Fig. 5.** The pipeline of iterative reconstruction procedure in OSDM.

## IV. EXPERIMENTS

### A. Data Specification

**AAPM Challenge Data:** The simulated data from human abdomen images provided by Mayo Clinic for the AAPM Low Dose CT Grand Challenge [43] are used for evaluation. The data includes high-dose CT scans from 10 patients, where 9 of them for training and the other one for evaluation. For fan-beam reconstruction network training, 5388 slices with 1 *mm* thickness and each of them covering $512\times512$ pixels are employed. In this study, only one projection data is used for OSMD training while 4700 projection data are used for U-Net and CNN training. There are 12 projection data picked as test set for all models. Artifact-free images, that are generated from normal-dose projection data by FBP algorithm, can be considered as ground truth. We extracted 1e5, 5e4, and 1e4 noise projection data from normal-dose projection for low-dose CT reconstruction. In fan-beam CT reconstruction, the Siddon's ray-driven algorithm [44, 45] is utilized to generate the projection data. The distance from rotation center to the source and detector are set to 40 *cm* respectively. The detector width is 41.3 *cm* including 720 detector elements and a total of projection views are evenly distributed over $360°$.

**CQ500 Dataset:** We test our method on Qure.ai's CQ500 dataset [46], which consists of 491 CT scans of human brains. We use 12 of those scans for testing. The remaining scans are omitted due to being outliers in terms of aspect ratio and number of slices. The chosen subset of the data has a resolution of $512\times512$ per slice and has between 101 and 645 slices per volume. The volumes are given in Hounsfield units in the range of [0, 4095] and normalized to [0, 1]. To get a similar level of noise, Poisson noise is added into the input training slices considering an exposure of $I_0 = 5\times10^5$ photons at each detector pixel before attenuation. The neural network training parameters and the reconstruction parameters are the same as those for the AAPM data.

### B. Model Training and Parameter Selection

In the experiments, our model is trained by the Adam algorithm with the learning rate $10^{-3}$ and Kaiming initialization is used to initialize the weights. The method is implemented in Python using Operator Discretization Library (ODL) [47] and PyTorch on a personal workstation with a GPU card (Tesla V100-PCIE-16GB). In the reconstruction stage, the iteration number is set to $N=1000$, $M=2$. Each time the prediction process of the outer loop is executed, the correction process of the inner loop is iterated twice by annealing Langevin. Referring to [34-37], $\eta$ in PWLS scheme is set as a constant 22000. The length of each gradient-descent step $\alpha$ in TV minimization is set to be 10. The singular value thresholding in SVD operation is 38 and the sliding window size is $8\times8$. Our source code can be publicly accessed at: https://github.com/yqx7150/OSDM.

### C. Quantitative Indices

To evaluate the quality of the reconstructed data, peak signal-to-noise ratio (PSNR), structural similarity index (SSIM), and mean squared error (MSE) are used for quantitative assessment.

PSNR describes the relationship of the maximum possible power of a signal with the power of noise corruption. Higher PSNR means better reconstruction quality. Denoting $I$ and $\tilde{I}$ to be the estimated reconstruction and ground-truth, PSNR is expressed as:

$$PSNR(I,\tilde{I}) = 20\log_{10}[\text{Max}(\tilde{I})/\|I-\tilde{I}\|_2] \quad (30)$$

The SSIM value is used to measure the similarity between

the ground-truth and reconstruction. SSIM is defined as:

$$SSIM(I,\tilde{I}) = \frac{(2\mu_I\mu_{\tilde{I}} + c_1)(2\sigma_{I\tilde{I}} + c_2)}{(\mu_I^2 + \mu_{\tilde{I}}^2 + c_1)(\sigma_I^2 + \sigma_{\tilde{I}}^2 + c_2)} \quad (31)$$

where $\mu_I$ and $\sigma_I^2$ are the average and variances of $I$. $\sigma_{I\tilde{I}}$ is the covariance of $I$ and $\tilde{I}$. $c_1$ and $c_2$ are used to maintain a stable constant. MSE is employed to evaluate the errors and it is defined as:

$$MSE(I,\tilde{I}) = \sum_{i=1}^{W} \|I_i - \tilde{I}_i\|_2 / W \quad (32)$$

where $W$ is the number of pixels within the reconstruction result. If MSE approaches to zero, the reconstructed image is closer to the reference image.

### D. Experimental Comparison

**Quick Comparison:** We compare our proposed method with four baseline techniques in low-dose CT reconstruction including FBP [2], SART-TV [48], CNN [49], and U-Net [50]. The involved parameters are set by the guidelines in their original papers.

For low-dose CT reconstruction of 1e5, 5e4, and 1e4 noise level, the PSNR, SSIM and MSE values of the reconstructed results from AAPM Challenge Dataset are listed in Table 1. The best PSNR and SSIM values of the reconstructed images with different projection dose are highlighted in bold. It is obvious that OSDM presents more details compared to the other methods. It can be observed that compared with other four methods, OSDM is able to achieve impressive average MSE value of 5.51e-5 and 7.74e-5 at 1e5 and 5e4 noise level in Table 1. The reconstructed images contain less artifacts and more details with decreased noise. It is exciting that the image reconstructed by OSDM method can reach 42.62 dB in the case of 1e5 noise. Thus, OSDM is able to achieve visible gains in terms of noise and artifacts suppression.

To further illustrate the merits of OSDM method, the reconstructed images and residual images are depicted in Figs. 6-7. As shown in Fig. 6, the FBP method results are the worst, as the performance of the analytical approach highly relies on noise. The results of SART-TV outperform those obtained by FBP algorithm. The sharper result with reasonable boundaries is obtained from the deep prior reconstruction approach. Nonetheless, reasonable geometrical details are missing in the results. Although CNN and U-Net algorithm achieves acceptable result, some edge details are still lost. On the contrary, the image reconstructed by OSDM method is very close to the ground truth with preserved details and structures.

TABLE I
RECONSTRUCTION PSNR/SSIM/MSEs OF AAPM CHALLENGE DATA USING DIFFERENT METHODS AT 1E5, 5E4 AND 1E4 NOISE LEVEL.

| Noise | FBP | SART-TV | CNN | U-Net | OSDM |
|---|---|---|---|---|---|
| 1e5 | 34.62/0.9252/3.66e-4 | 41.03/0.9892/8.65e-5 | 41.36/0.9904/8.20e-5 | 42.52/**0.9921**/6.28e-5 | **42.62**/0.9899/**5.51e-5** |
| 5e4 | 32.43/0.8866/5.81e-4 | 38.72/0.9786/1.39e-4 | 39.26/0.9872/1.28e-4 | **41.34**/**0.9897**/7.83e-5 | 41.20/0.9857/**7.74e-5** |
| 1e4 | 25.78/0.6897/2.69e-3 | 29.58/0.8710/1.18e-3 | 37.58/**0.9795**/1.81e-4 | **38.50**/0.9782/**1.44e-4** | 37.43/0.9683/1.83e-4 |

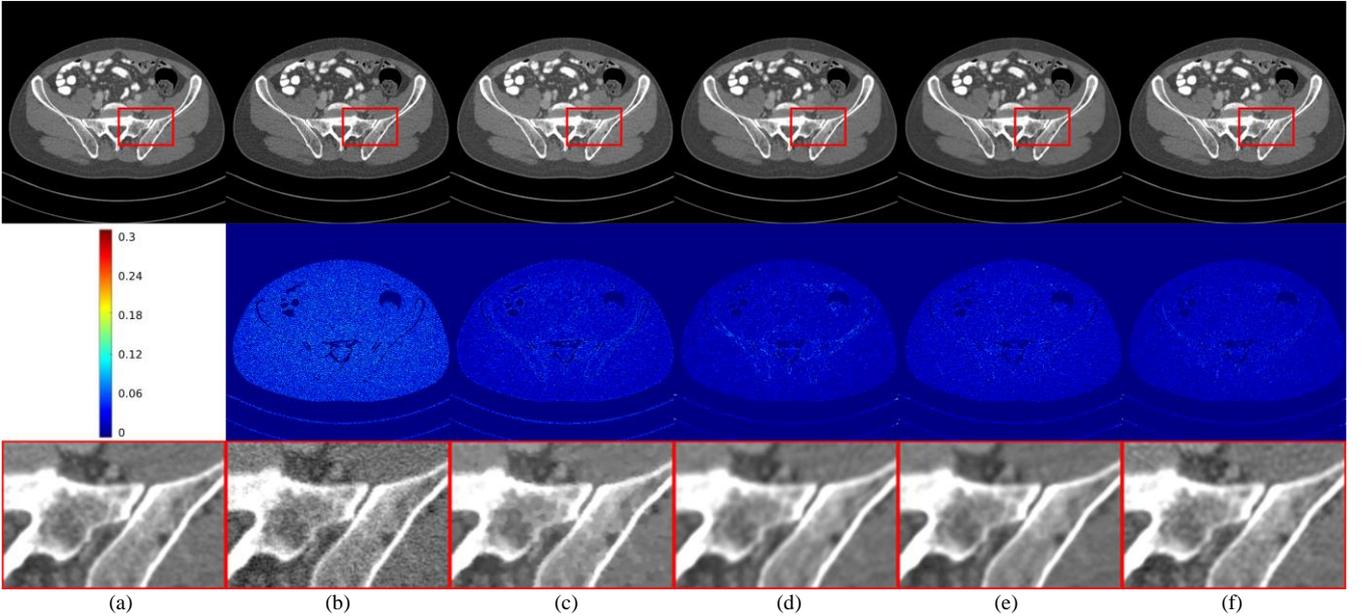

**Fig. 6.** Reconstruction results from 1e5 noise using different methods. (a) The reference image versus the images reconstructed by (b) FBP, (c) SART-TV, (d) CNN, (e) U-Net, and (f) OSDM. The display windows are [-250,600]. The second row shows residuals between the reference and reconstructed images.

Compared with the results reconstructed from 1e4 noise, the results reconstructed from 1e5 noise exhibit obvious improvements of image quality as indicated in Figs. 6-7. The image reconstructed by FBP algorithm not only contains artifacts, but also some important structural features are not well preserved. Meanwhile, images reconstructed by U-Net still suffer from severe steaking artifacts. Moreover, CNN can distinguish some details, but the edges are over-smoothed. Finally, the image reconstructed by OSDM method preserves more structural details while suppressing streaking artifacts.

Since the unsupervised model is employed, there is no need to train the model separately for different noise level situations. In the meanwhile, mere one data of full-sampled is required to train our model sufficiently. Therefore, in the case of high noise, our method performs flat compared with other baseline methods. Notably, our approach shows promising results in the case of low noise.

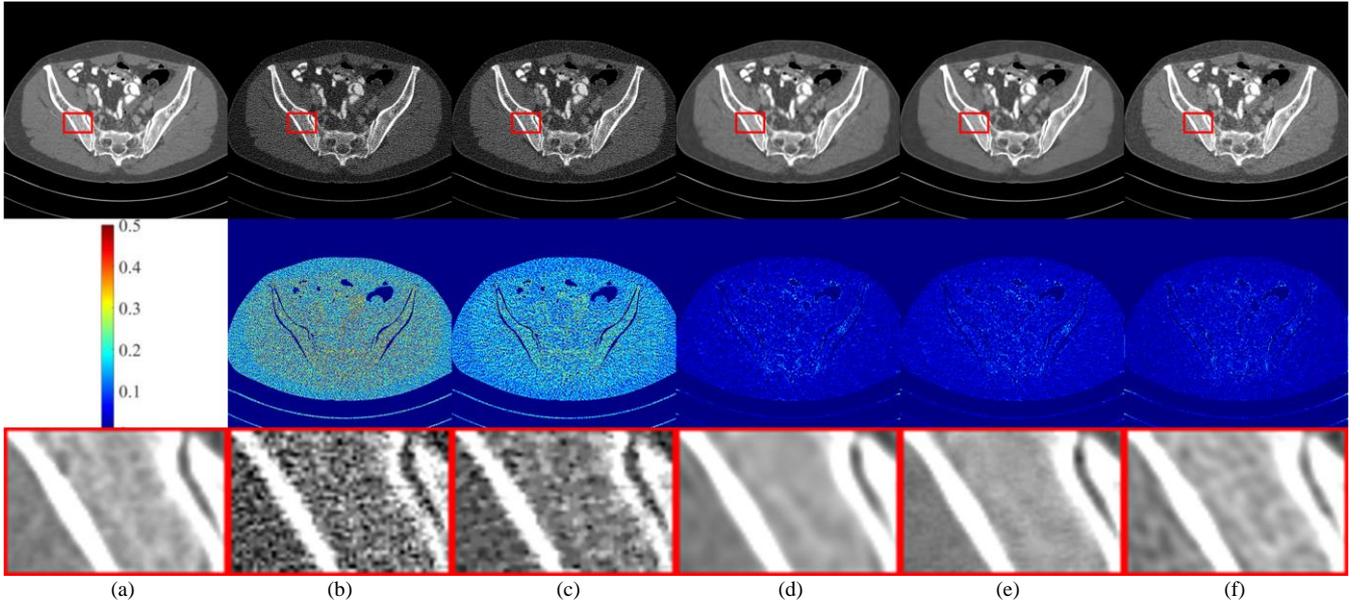

(a)     (b)     (c)     (d)     (e)     (f)

Fig. 7. Reconstruction results from 1e4 noise using different methods. (a) The reference image versus the images reconstructed by (b) FBP, (c) SART-TV, (d) CNN, (e) U-Net, and (f) OSDM. The display windows are [-250,600]. The second row shows residuals between the reference and reconstructed images.

*Generalization and Robustness Test:* To compare the generalization of CNN, U-Net and OSDM, we no longer pretrain the model separately at each noise level data for CNN and U-Net. In this experiment, the model trained at 1e5, 5e4, and 1e4 noise level are all used for reconstructing data under 1e5 noise scale. Two supervised methods CNN and U-Net are chosen as baseline. Table II presents all the results, and the best value of each metric is marked in black bold. Obviously, our method scores the highest PSNR, SSIM, and MSE in most occasions. OSDM is an unsupervised method which only need one training model.

It's impossible to train every noise level in real life, which means CNN, U-Net or any other supervised methods are unable to apply for different noises or dose. From Table II, we can discover that when the training noise is changing, the results of CNN and U-Net are getting worse quickly.

TABLE II
RECONSTRUCTION PSNR/SSIM/MSES OF AAPM CHALLENGE DATA USING DIFFERENT TRAINING MODEL AT 1E5 NOISE LEVEL.

| Training Model | CNN | U-Net | OSDM |
|---|---|---|---|
| 1e5 | 41.36/0.9904/ 8.20e-5 | 42.52/**0.9921**/ 6.28e-5 | **42.62**/0.9899/ **5.51e-5** |
| 5e4 | 41.00/0.9898/ 1.10e-4 | 41.97/**0.9907**/ 7.54e-5 | **42.62**/0.9899/ **5.51e-5** |
| 1e4 | 36.40/0.9830/ 2.87e-4 | 40.15/0.9868/ 1.11e-4 | **42.62**/**0.9899**/ **5.51e-5** |

For further proving the generalization of the proposed unsupervised learning scheme, OSDM is migrated to CQ500 dataset at 1e4 noise level qualitatively and quantitatively. Especially, we learn the prior knowledge on AAPM challenge data and test our model on CQ500 dataset. Table III records all the results, and the value of each metric maintains well. Intuitively, our method scores the highest PSNR and MSE of 38.15 dB and 1.55e-4 on CQ500 dataset compared with other methods. At the same time, good image quality evaluation metrics are obtained when testing on CQ500 dataset. Therefore, it implies that OSDM has great adaptability on different dataset.

To visually illustrate the performance of OSDM, we perform qualitative comparisons of different test datasets. In addition, for better evaluation of image quality, Fig. 8 depicts the zoomed regions-of-interest as marked by the red rectangles. From the results in Fig. 8(c), we can see that our method achieves the best performance in terms of noise-artifact suppression and tissue feature preservation. Meanwhile, the results reconstructed by FBP are in bad shape. The results reconstructed by CNN and U-Net lead to a few noises and artifacts compared with those of our method. The experiment proves exactly the good generalization of OSDM.

TABLE III
RECONSTRUCTION PSNR/SSIM/MSES OF CQ500 DATASET USING DIFFERENT METHODS AT 1E4 NOISE LEVEL.

| 1e4 | FBP | CNN | U-Net | OSDM |
|---|---|---|---|---|
| PSNR | 19.09 | 36.65 | 32.56 | **38.15** |
| SSIM | 0.5269 | **0.9836** | 0.9545 | 0.9818 |
| MSE | 1.35e-2 | 2.25e-4 | 5.68e-4 | **1.55e-4** |

TABLE IV
RECONSTRUCTION PSNR/SSIM/MSES OF CQ500 DATASET USING DIFFERENT METHODS AT 1E3 NOISE LEVEL.

| 1e3 | FBP | CNN | U-Net | OSDM |
|---|---|---|---|---|
| PSNR | 11.04 | 18.16 | 18.38 | **33.65** |
| SSIM | 0.1879 | 0.6219 | 0.5817 | **0.9608** |
| MSE | 7.95e-2 | 1.54e-2 | 1.46e-2 | **4.38e-4** |

To explore the generalization of OSDM better, based on testing with CQ500 dataset, the pretrained model at 1e4 noise level is used to reconstruct the data at 1e3 noise level for CNN and U-Net while OSMD still use the old one. Quantitative results reconstructed from different reconstruction methods are tabulated in Table IV. It can be observed that OSDM outperforms the other methods in a trend similar to what we have seen from the reconstruction images and produces the highest PSNR. After changing the test data and noise level, OSDM is way better than CNN and U-Net.

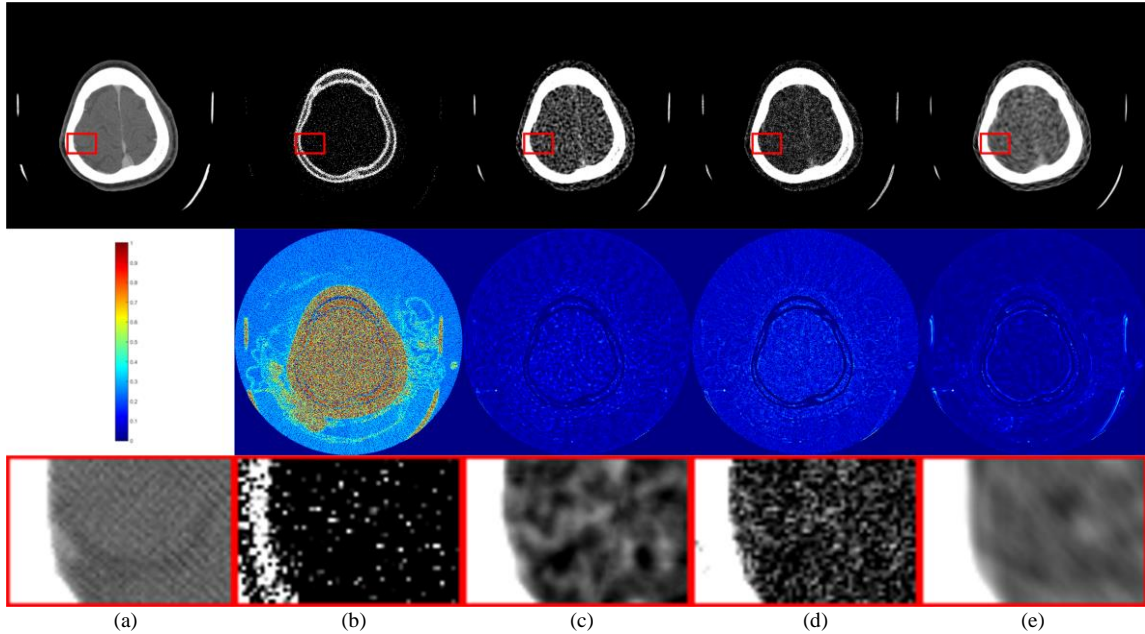

(a) (b) (c) (d) (e)

**Fig. 8.** Reconstruction results from 1e4 noise using different methods. (a) The reference image versus the images reconstructed by (b) FBP, (c) CNN, (d) U-Net, (e) OSDM. The display windows are [350,550]. The second row shows residuals between the reference and reconstructed images.

### E. Ablation Study

**Different Components in OSDM:** In order to verify the necessity of corresponding modules in our method, OSDM is compared qualitatively and quantitatively with Hankel based singular value decomposition (H-SVD) [51] and H-SVD with diffusion model at 1e5, 5e4, and 1e4 noise level. The above methods all regard PWLS as the data fidelity item to improve the quality of image reconstruction. It is worth noting that our method employs the TV minimization technique embedding into diffusion model to suppress the noise generation in the reconstruction process. In addition, H-SVD is just a traditional algorithm without network. We calculate the average value of the test set and present the results in Table V to verify the effectiveness of OSDM. Table V lists the quantitative results for the reconstructions from 1e5, 5e4, and 1e4 noise projection dose, where the best results are highlighted in bold. It can be observed that OSDM method yields the best results in terms of PSNR, SSIM, and MSE values, which are consistent with the visual effects. Specifically, compared to H-SVD method, OSDM method has a notable PSNR gain of 0.97 dB, 0.96 dB, and 1.73 dB in cases of 1e5, 5e4, and 1e4 noise projection dose. It is worth noting that our results are also impressive in terms of SSIM and MSE.

TABLE V
RECONSTRUCTION PSNR/SSIM/MSEs OF AAPM CHALLENGE DATA USING DIFFERENT METHODS AT 1E5, 5E4 AND 1E4 NOISE LEVEL.

| Noise | H-SVD | H-SVD with Diffusion Model | OSDM |
|---|---|---|---|
| 1e5 | 41.65±0.93/0.9883±0.0014/6.99e-5 | 42.02±0.86/0.9884±0.0014/6.40e-5 | **42.62±0.51/0.9899±0.0011/5.51e-5** |
| 5e4 | 40.24±1.12/0.9839±0.0024/9.80e-5 | 40.79±0.60/0.9841±0.0019/8.40e-4 | **41.20±0.93/0.9857±0.0026/7.74e-5** |
| 1e4 | 35.70±0.72/0.9551±0.0050/2.73e-4 | 36.01±0.76/0.9615±0.0053/2.54e-4 | **37.43±0.78/0.9683±0.0062/1.83e-4** |

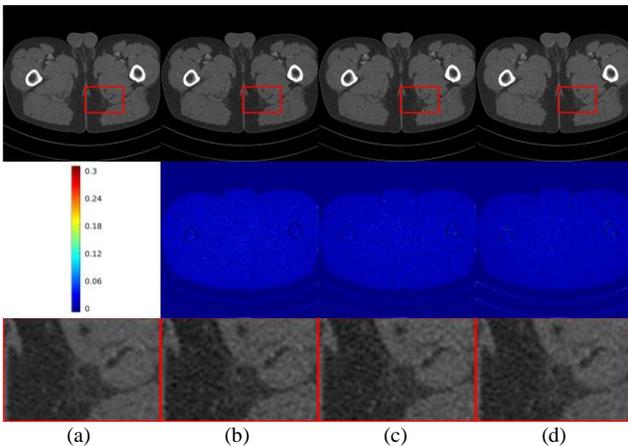

(a) (b) (c) (d)

**Fig. 9.** Reconstruction results from 1e5 noise using different methods. (a) The reference image versus the images reconstructed by (b) H-SVD, (c) H-SVD with Diffusion Model, (d) OSDM. The display windows are [-250,600]. The second row shows residuals between the reference and reconstructed images.

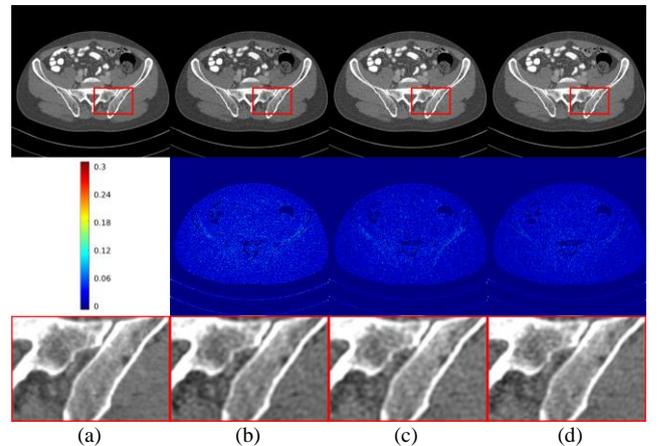

(a) (b) (c) (d)

**Fig. 10.** Reconstruction results from 5e4 noise using different methods. (a) The reference image versus the images reconstructed by (b) H-SVD, (c) H-SVD with Diffusion Model, (d) OSDM. The display windows are [-250,600]. The second row shows residuals between the reference and reconstructed images.

In Figs. 9-10, the reconstruction results of three methods are presented clearly. The quality of H-SVD image is relatively poor that some details and important structures cannot be distinguished. H-SVD with diffusion model can remove some artifacts while some important details are lost and the edges of the results are blurred as indicated in Fig. 9. On the contrary, OSDM results in Fig. 9 can achieve the best performance in terms of noise suppression and structural detail preservation. It can be seen from the residual images that the proposed method provides the best performance as shown in Fig. 9.

Fig. 10 depicts the reconstructed images from 5e4 noise with different methods. Compared with the images in Fig. 9, the artifacts in the image reconstructed by all algorithm are increased as shown in Fig. 10. The H-SVD compromises structural details and suffers from notorious blocky artifacts. Some details in H-SVD with diffusion model are over-smoothed. Excitingly, OSDM can retain more structure information and details than other methods. From the ablation study above, we can easily conclude that each module, including H-SVD, diffusion model and TV minimization technique, plays an important role in OSDM.

*One Sample VS Few Samples:* By changing the number of samples, the feasibility of one sample is verified as following. Here, we choose 1, 50, and 100 samples as training set. Table VI presents all the results, and the best value of each metric is marked in black bold. For a more detailed comparison of PSNR value, three decimal places of PSNR are kept deliberately. Although all data metrics are not optimal when using only one sample for training, there is no big gap between them. From Table VI, we could observe that as the number of training samples reach 50, every metrics begin to stabilize. Therefore, employing few samples or even one sample is exactly feasible.

TABLE VI
RECONSTRUCTION PSNR/SSIM/MSEs OF AAPM CHALLENGE DATA USING DIFFERENT NUMBER OF SAMPLES AT 1E5 NOISE LEVEL.

| 1e5 | 1 | 50 | 100 |
|---|---|---|---|
| PSNR | 42.616 | 42.950 | **42.953** |
| SSIM | 0.9899 | 0.9899 | **0.9903** |
| MSE | 5.51e-5 | **5.13e-5** | 5.14e-5 |

## V. CONCLUSIONS

Although the deep learning-based CT reconstruction methods have achieved great successes in the past several years, the generalizability and robustness of the trained networks was still an open problem. In this work, we presented a score-based diffusion model for low-dose CT reconstruction, where the PC sampling was used to generate the sinogram. Specifically, we used a fully unsupervised technique to train a score-based diffusion model to capture the prior distribution of sinogram data. Then, at the iterative inference stage, the numerical SDE solver, PWLS strategy, TV minimization, and LR operation were performed alternatively to achieve image reconstruction. More precisely, the predictor referred to a numerical solver for the reverse-time SDE and Langevin dynamics was considered as the corrector. The effectiveness and generalizability of OSDM were verified on AAPM challenge dataset and CQ500 dataset. Experimental results illustrated that the proposed method could effectively suppress the streaking artifact and preserving important sinogram details for low-dose CT reconstruction. Furthermore, due to patient privacy, data security and other issues, it was difficult to conduct data collection in medical fields. In this circumstance, one sample or few samples based CT reconstruction was crucial. Thus, we have proposed OSDM method which could reach an adequate effect by one sample and perform superior with few samples.

REFERENCES

[1] M. Bakator and D. Radosav, "Deep learning and medical diagnosis: A review of literature," *Multimodal Tech. Interact.,* vol. 2, no. 3, pp. 47 Aug. 2018.
[2] D. J. Brenner and E. J. Hall, "Computed tomography-an increasing source of radiation exposure," *New Engl. J. Med.,* vol. 357, no. 22, pp. 2277-2284, 2007.
[3] E. Y. Sidky and X. Pan, "Image reconstruction in circular cone-beam computed tomography by constrained, total-variation minimization," *Phys. Med. Biol.,* vol. 53, no. 17, pp. 4777, Aug. 2008.
[4] H. Yu and G. Wang, "Compressed sensing based interior tomography," *Phys. Med. Biol.,* vol. 54, no. 9, pp. 2791-2805, 2009.
[5] Y. Chen, D. Gao, C. Nie, L. Luo, W. Chen, X. Yin, and Y. Lin, "Bayesian statistical reconstruction for low-dose x-ray computed tomography using an adaptive weighting nonlocal prior," *Comput. Med. Imag. Graph.,* vol. 33, no. 7, pp. 495-500, 2009.
[6] J. Liu, Y. Hu, J. Yang, Y. Chen, and H. Shu, L. Luo, Q. Feng, Z. Gui, G. Coatrieux, "3D feature constrained reconstruction for low dose CT Imaging," *IEEE Trans. Circuits Syst. Video Technol.,* vol. 28, no. 5, pp. 1232-1247, 2017.
[7] X. Zheng, X. Lu, S. Ravishankar, Y. Long, and J. Fessler, "Low dose CT image reconstruction with learned sparsifying transform," *in Proc. IEEE Workshop on Image, Video, Multidim. Signal Proc.,* pp. 1-5, 2016.
[8] W. Zheng, B. Yang, Y. Xiao, J. Tian, S. Liu, and L. Yin, "Low-dose CT image post-processing based on learn-type sparse transform," *Sensors,* vol. 22, no. 8, pp. 2883, 2022.
[9] Y. Chen, X. Yin, L. Shi, H. Shu, L. Luo, J.-L. Coatrieux, and C. Toumoulin, "Improving abdomen tumor low-dose CT images using a fast dictionary learning based processing," *Phys. Med. Biol.* vol. 58, no. 16, pp. 5803–5820, 2013.
[10] X. Yin, Q. Zhao, J. Liu, W. Yang, J. Yang, G. Quan, Y. Chen, H. Shu, L. Luo, and J.-L. Coatrieux, "Domain progressive 3D residual convolution network to improve low-dose CT imaging," *IEEE Trans. Med. Imag.,* vol. 38, no. 12, pp. 2903-2913, 2019.
[11] T. Humphries, D. Si, S. Coulter, M. Simms, and R. Xing, "Comparison of deep learning approaches to low dose CT using low intensity and sparse view data," *In Med. Imag. 2019: Phys. of Med. Imag.,* vol. 10948, pp. 1048-1054, Mar. 2019.
[12] M. U. Ghani, and W. C. Karl, "CNN based sinogram denoising for low-dose CT," *In Math. in Imag.,* pp. MM2D-5, Jun. 2018.
[13] M. Beister, D. Kolditz, and W. A. Kalender, "Iterative reconstruction methods in X-ray CT," *Phys. Med,* vol. 28, no. 2, pp. 94–108, Apr. 2012.
[14] S. Ramani and J. A. Fessler, "A splitting-based iterative algorithm for accelerated statistical X-ray CT reconstruction," *IEEE Trans. Med. Imag.,* vol. 31, no. 3, pp. 677–688, Mar. 2012.
[15] Y. Liu, J. Kang, Z. Li, Q. Zhang, and Z. Gui, "Low-dose CT noise reduction based on local total variation and improved wavelet residual CNN," *J. of X-Ray Sci. and Technol.,* vol. Pre-press, no. Pre-press, pp. 1-14, 2022.
[16] X. Deng, Y. Zhao, and H. Li, "Projection data smoothing through noise-level weighted total variation regularization for low-dose computed tomography," *J. of X-ray Sci. and Technol.,* vol. 27, no. 3, pp. 537-557, 2019.
[17] S. V. M. Sagheer, and S. N. George, "Denoising of low-dose CT images via low-rank tensor modeling and total variation regularization," *Artif. Intel. in Med.,* vol. 94, pp. 1-17, 2019.
[18] J. Sohl-Dickstein, E. Weiss, N. Maheswaranathan, and S. Ganguli. "Deep unsupervised learning using nonequilibrium thermodynamics," *In Int. Conf. on Mach. Learn.,* vol. 37, pp. 2256–2265, 2015.
[19] L. Yang, Z. Zhang, and S. Hong. "Diffusion models: A comprehensive survey of methods and applications," *arXiv preprint arXiv:2209.00796,* 2022.
[20] Q. Gao, and H. Shan. "CoCoDiff: a contextual conditional diffusion model for low-dose CT image denoising," *In Developments in X-Ray Tomography XIV,* vol. 12242, Oct. 2022.
[21] Q. Lyu and G. Wang. "Conversion Between CT and MRI Images

Using Diffusion and Score-Matching Models," *arXiv preprint arXiv:2209.12104*, 2022

[22] Y. Song, J. Sohl-Dickstein, D. P. Kingma, A. Kumar, S. Ermon, and B. Poole. "Score-based generative modeling through stochastic differential equations," *arXiv preprint arXiv:2011.13456*, 2020.

[23] H. Chung, and J. C. Yea, "Score-based diffusion models for accelerated MRI," *arXiv preprint arXiv:2110.05243*, 2021.

[24] A. Jalal, M. Arvinte, Daras G, E. Price, A. Dimakis, and J. Tamir, "Robust compressed sensing MRI with deep generative priors," *Adv. Neural Inf. Process. Syst.*, vol. 34, 2021.

[25] Y. Song, L. Shen, L. Xing, and S. Ermon. "Solving inverse problems in medical imaging with score-based generative models," *arXiv preprint arXiv:2111.08005,2021*

[26] K. H. Jin and J. C. Ye, "Sparse and low-rank decomposition of a hankel structured matrix for impulse noise removal," *IEEE Trans. Image Process*, vol. 27, no. 3, pp. 1448–1461, 2018.

[27] K. H. Jin, J.-Y. Um, D. Lee, J. Lee, S.-H. Park, and J. C. Ye, "MRI artifact correction using sparse+low-rank decomposition of annihilating filter-based Hankel matrix," *Magn. Reson. Med*, vol. 78, no. 1, pp. 327–340, Jul. 2017.

[28] J. Min, L. Carlini, M. Unser, S. Manley, and J. C. Ye, "Fast live cell imaging at nanometer scale using annihilating filter-based low-rank Hankel matrix approach," *Proc. SPIE*, vol. 9597, pp. 204-211, Sep. 2015.

[29] G. Wang, J. C. Ye, K. Mueller, and J. A. Fessler, "Image reconstruction is a new frontier of machine learning," *IEEE Trans. Med. Imag.*, vol. 37, no. 6, pp. 1289-1296, 2018.

[30] T. R. Shaham, T. Dekel, and T. Michaeli, "Singan: learning a generative model from a single natural image," *Proc. IEEE Int. Conf. Comput. Vis.*, pp. 4570-4580, 2019.

[31] V. Sushko, J. Gall, and A. Khoreva, "One-shot gan: Learning to generate samples from single images and videos," *Proc. IEEE Conf. Comput. Vis. Pattern Recognit.*, pp. 2596-2600, 2021.

[32] Z. Zheng, J. Xie, and P. Li, "Patchwise generative convnet: Training energy-based models from a single natural image for internal learning," *Proc. IEEE Conf. Comput. Vis. Pattern Recognit.*, pp. 2961-2970, 2021.

[33] X. Jia, Y. Lou, R. Li, W. Y. Song and S. B. Jiang. "GPU-based fast cone beam CT reconstruction from undersampled and noisy projection data via total variation," *Med. Phys.*, vol. 37, no. 4, pp. 1757-1760, 2010.

[34] T. Li, X. Li, J. Wang, J. Wen, H. Lu, J. Hsieh, and Z. Liang, "Nonlinear sinogram smoothing for low-dose X-ray CT," *IEEE Trans. on Nuclear Sci.,* vol. 51, no. 5, pp. 2505-2513, 2004.

[35] J. Wang, T. Li, H. Lu, and Z. Liang, "Penalized weighted least-squares approach to sinogram noise reduction and image reconstruction for low-dose X-ray computed tomography," *IEEE Trans. Med. Imag.*, vol. 25, no. 10, pp. 1272–1283, 2006.

[36] J. Wang, H. Lu, J. Wen, and Z. Liang, "Multiscale penalized weighted least-squares sinogram restoration for low-dose X-ray computed tomography," *IEEE Trans. on Biomed. Eng.*, vol. 55, no. 3, pp. 1022-1031, 2008.

[37] J. Wang, T. Li, and L. Xing, "Iterative image reconstruction for CBCT using edge-preserving priori," *Med. Phys.*, vol. 36, no. 1, pp. 252-260, 2009.

[38] Y. Zhang, J. Zhang, and H. Lu, "Statistical sinogram smoothing for low-dose CT with segmentation-based adaptive filtering," *IEEE Trans. on Nuclear Sci.*, vol. 57, no. 5, pp. 2587-2598, 2010.

[39] E. Kang, J. Min, and J. C. Ye, "A deep convolutional neural network using directional wavelets for low-dose X-ray CT reconstruction," *Medical Physics*, vol. 44, no. 10, pp. e360-e375, 2017.

[40] O. Ronneberger, P. Fischer, and T. Brox, "U-net: Convolutional networks for biomedical image segmentation," Int. Conf. on *Med. Image Comput. Comput.-Assisted Intervention*, pp. 234-241, 2015.

[41] Y. Song and S. Ermon, "Generative modeling by estimating gradients of the data distribution," *in Adv. Neural Inf. Process. Syst.*, pp. 11918-11930, 2019.

[42] I. A. Elbakri and J.A. Fessler, "Statistical image reconstruction for polyenergetic x-ray computed tomography," *IEEE Trans. Med. Imag.*, vol. 21, no. 2, pp. 89-99, 2002.

[43] Low Dose CT Grand Challenge. Accessed: Apr. 6, 2017. [Online]. Available: http://www.aapm.org/GrandChallenge/LowDoseCT/.

[44] R. L. Siddon, "Fast calculation of the exact radiological path fora three-dimensional CT array," *Med. Phys.*, vol. 12, no. 2, pp. 252–255, 1985.

[45] F. Jacobs, E. Sundermann, B. De Sutter, M. Christiaens, and I. Lemahieu, "A fast algorithm to calculate the exact radio logical path through a pixel or voxel space," *J. of Comput. and Inf. Technol.*, vol. 6, no. 1, pp. 89–94, 1998.

[46] S. Chilamkurthy R. Ghosh, S. Tanamala, M. Biviji, N. G. Campeau, V. K. Venugopal, V. Mahajan, P. Rao, and P. Warier. "Development and validation of deep learning algorithms for detection of critical findings in head CT scans," *arXiv preprint arXiv:1803.05854*, 2018.

[47] J. Adler, H. Kohr, and O. Oktem, "Operator discretization library (ODL)," Software available from https://github.com/odlgroup/odl, vol. 5, 2017.

[48] Y. Censor, and T. Elfving, "Block-iterative algorithms with diagonally scaled oblique projections for the linear feasibility problem," *SIAM J. Matrix Anal. Appl*, vol. 24, no. 1, pp. 40-58, 2002.

[49] K. Zhang, W. Zuo, Y. Chen, D. Meng, and L. Zhang. "Beyond a gaussian denoiser: Residual learning of deep cnn for image denoising," *IEEE Trans. on Image Process.*, vol. 26, no. 7, pp. 3142-3155, 2017.

[50] H. Lee, J. Lee, H. Kim, B. Cho, and S. Cho, "Deep-neural-network-based sinogram synthesis for sparse-view CT image reconstruction," *IEEE Trans. on Radiat. and Plasma Med. Sci.*, vol. 3, no. 2, pp. 109-119, 2019.

[51] P. J. Shin, P. E. Larson, M. A. Ohliger, M. Elad, J. M. Pauly, D. B. Vigneron, and M. Lustig. "Calibrationless parallel imaging reconstruction based on structured low-rank matrix completion," *Magn. Reson. in Med.*, vol. 72, no. 4, pp. 959-970, 2014.